# Golden Sections of Interatomic Distances as Exact Ionic Radii and Additivity of Atomic and Ionic Radii in Chemical Bonds

Raji Heyrovska

Institute of Biophysics, Academy of Sciences of the Czech Republic;

Email: rheyrovs@hotmail.com;

## Abstract

The Golden ratio which appears in the geometry of a variety of creations in Nature is found to arise right in the Bohr radius of the hydrogen atom due to the opposite charges of the electron and proton. The bond length of the hydrogen molecule is the diagonal of a square on the Bohr radius and hence also has two Golden sections, which form the cationic and anionic radii of hydrogen. It is shown here that these radii account for the bond lengths of many hydrides when added to the atomic and Golden ratio based ionic radii of many other atoms.

## 1. Introduction

The covalent [1] or bonding atomic radius [2], d(A) of an atom A is defined as half the bond length d(AA), and the covalent bond length d(AB) between two different atoms A and B is [1,2] the sum d(A) + d(B) as shown:

$$d(A) = d(AA)/2; \; d(B) = d(BB)/2; \; d(AB) = d(A) + d(B) \qquad (1a,b,c)$$

Many observed bond lengths have been shown to be the sums of the covalent radii of the adjacent atoms [1,2]. By using the appropriate covalent radii, it has been shown recently that the known inter-atomic distances in molecules like nucleic acids [3a,b], caffeine and related molecules [4], amino acids [5], graphene, benzene and methane [6] are all sums of the radii of the adjacent atoms.

For partially and completely ionic bonds, the atomic and ionic radii, as the case may be, are also additive, where the ions have the Golden ratio based radii [7,8]. The latter ionic radii are described here after a brief introduction to the Golden ratio [9a-c].

The Golden ratio ($\phi$) is the ratio a/b of two numbers a and b which are such that

$$a/b = (a + b)/a \qquad (2a)$$

$$(1/b) - (1/a) = 1/(a + b) \qquad (2b)$$

From the above one obtains Eq. 3 for the ratio (a/b),

$$(a/b)^2 = (a/b) + 1 \qquad (3)$$

$$a/b = (1 + 5^{1/2})/2 = 1.618.. = \phi = (a + b)/a; \qquad (4)$$

$$\phi^2 = \phi + 1 = (a + b)/b; (1/\phi^2) + (1/\phi) = 1 \qquad (5a,b)$$

$$1/\phi = 0.618..; (1/\phi^2) = 0.382..; (1/\phi) - (1/\phi^2) = 1/\phi^3 = 0.236.. \qquad (6a,b,c)$$

Eq. 4a gives the Golden ratio $\phi$ as the positive root for a/b, and since it involves $5^{1/2}$ = 2.236…, the decimal places in 1.618… are numerous. The two sections a and b are called the Golden sections of their sum (a + b), and an equation of the form of Eq. 3 is called the Golden quadratic. Eq. 5a represents the Golden quadratic in terms of $\phi$. Eq.

5b shows the two Golden sections of unity, Eqs. 6a,b,c give the values of the Golden sections of unity and their difference.

Further, $\phi/2 = \cos 36^0$ and $2\sin 18^0 = 1/\phi$, as shown in Eqs. 7a,b, are exact trigonometric ratios. The angles $18^0$, $36^0$, $54^0$ and $72^0$ appear extensively in the regular pentagon, pentagram and decagon [9a-c]. See Figs. 1a-c.

$$\cos 36^0 = \sin 54^0 = (1 + 5^{1/2})/4 = \phi/2 = 0.809.. \qquad (7a)$$

$$\sin 36^0/\cos 18^0 = 2\sin 18^0 = 2\cos 72^0 = 2/(1 + 5^{1/2}) = 1/\phi = 0.618.. \qquad (7b)$$

$\phi$ and the Fibonacci numbers (where each term is the sum of the previous two) are closely related since the ratio of any Fibonacci number to its previous one oscillates around $\phi$ and finally tends to it when the numbers are large. $\phi$ itself forms a 'Golden Fibonacci series' which at the same time is also a geometric series [10],

$$\ldots, 1/\phi^2, 1/\phi, 1, \phi, \phi^2, \phi^3, \ldots.. \qquad (8)$$

The Golden ratio is also called the Divine ratio [9,11] since it appears in the geometry of a wide variety of Nature's spontaneous creations.

Any given distance AB can be divided into two Golden sections, AP and BP by locating P, the Golden point. In Fig. 1a, [9b] the line BC = AB/2 is drawn perpendicular to AB and the points D and P are marked on AC and AB respectively such that BC = CD and AP = AD. In Fig. 1b, [9c] AB is the radius of a circle (which circumscribes a pentagon and a hexagon and inscribes a square. AE = AB/2 and EF = $(5^{1/2}/2)$AB. The point P is marked on AB such that EP = EF. The Golden point P here corresponds to the point D in [9c]. Fig. 1c shows the pentagon, the pentagram and decagon, where the Golden ratio occurs extensively. All the details and the Golden ratios are given in the boxes to the right of the figures.

## 2. THE GOLDEN RATIO IN ATOMIC PHYSICS: THE BOHR RADIUS

The energy needed to ionize a hydrogen atom is the energy necessary to pull apart the oppositely charged proton and electron, $p^+$ and $e^-$ respectively, against their coulombic attraction. Hence the ionization potential $I_H = (e/\kappa a_B)$ is the difference (or the algebraic sum) of the potentials $I_p$ (= $e/\kappa a_{B,p}$) and $I_e$ (= $-e/\kappa a_{B,e}$) at ionization of $p^+$ and $e^-$, where $a_B = a_{B,p} + a_{B,e}$ is the Bohr radius (= distance between $p^+$ and $e^-$) and $\kappa$ is the electric permittivity. This gives the relations [7],

$$I_H = (e/\kappa a_B) = I_p + I_e = (e/\kappa)[(1/a_{B,p}) - (1/a_{B,e})] \quad (9a)$$

$$(1/a_B) = (1/a_{B,p}) - (1/a_{B,e});\ a_B = a_{B,p} + a_{B,e};\ (a_{B,e}/a_{B,p})^2 = (a_{B,e}/a_{B,p}) - 1 = 0 \quad (9b,c,d)$$

$$a_{B,e}/a_{B,p} = \phi = (1 + 5^{1/2})/2 = 1.618..\ ;\ a_{B,p} = (a_B/\phi^2) \text{ and } a_{B,e} = (a_B/\phi) \quad (10a,b,c)$$

On combining Eqs. 9b,c, one gets the Golden quadratic Eq. 9d, the positive root of which is $\phi$, as given by Eq. 10a; $a_{B,p}$ = 0.20 Å and $a_{B,e}$ = 0.33 Å are the Golden sections of $a_B$ (= 0.529 Å [12]) as shown in Eqs. 10b,c. See Fig. 2.

The de Broglie wavelength, $\lambda_{dB} = 2\pi a_B$ is the circumference of the Bohr circle. Since $a_B$ consists of two Golden sections, the de Broglie wavelength is also the sum of the circumferences of two Golden circles: $\lambda_{dB,e} = 2\pi a_{B,e}$ and $\lambda_{dB,p} = 2\pi a_{B,p}$, with radii $a_{B,e}$ and $a_{B,p}$ respectively. Alternately, $\lambda_{dB,e}$ and $\lambda_{dB,p}$ can be considered as two sections of the Bohr circle with radius $a_B$, corresponding to the Golden angles, $360/\phi$ (= $222.49^0$) and $360/\phi^2$ (= $137.51^0$).

It is found [10] that the fine structure constant ($\alpha = \lambda_{CH}/\lambda_{dB} = 1/137.036$), Compton wavelength ($\lambda_{CH}$), relativity factor ($\gamma$), the contribution $\lambda_{CHi} = \phi 2\pi r_{\mu,H}$ from the sum of the intrinsic radii of the electron and proton ($r_{\mu,H}$, calculated from the magnetic moment anomalies), are all related as follows:

$$\alpha - (1 - \gamma)/\gamma = \phi^2/360; \;\; \gamma = \lambda_{dB}/(\lambda_{dB} + \lambda_{C,H,i}) = 0.99997(5) \qquad (11a,b)$$

The distances, $\lambda_{C,H}$, ($\lambda_{C,H} - \lambda_{C,H,i}$) and $\lambda_{C,H,i}$ correspond to small arc lengths on the Bohr circle of circumference $\lambda_{dB}$, subtended by central angles of $2.627^0$, $2.618^0$ (= $\phi^2$) and $0.009^0$ (= 2.627 - 2.618) respectively. The angle $0.009^0 = 360(1 - \gamma)/\gamma = 0.009(6)^0$, is the advance of the perihelion in Sommerfeld's theory of the hydrogen atom.

**3. GOLDEN SECTIONS OF THE INTERATOMIC DISTANCE IN $H_2$:**

The inter-atomic distance in the simplest diatomic molecule, $H_2$ is the covalent bond length d(HH) = 2d(H) = 0.74 Å [1], 0.749 Å [13] where d(H) = 0.37 Å is the covalent atomic radius [$R_{cov}$ = d(H), semi-covalent bond distance]. It is the diagonal of a square with $a_B$ (= 0.529Å [12]) as a side, with the two electrons and protons at the opposite corners of the diagonal. Since $a_B$ has two Golden sections (Eqs. 10b,c), one finds that

$$d(HH) = 2^{1/2}a_B = 2^{1/2}(a_{B,p} + a_{B,e}) = d(HH)/\phi^2 + d(HH)/\phi = d(H^+) + d(H^-) \qquad (12)$$

where $d(H^+) = d(HH)/\phi^2 = 2^{1/2}(a_{B,p})$ = 0.28 Å and $d(H^-) = d(HH)/\phi = 2^{1/2}(a_{B,e})$ = 0.46 Å are the Golden ratio based cationic and anionic radii of H [7]. See Fig. 3.

## 4. THE GOLDEN RATIO BASED IONIC RADII OF HYDROGEN AND BOND LENGTHS OF HYDRIDES

The value 0.28 Å suggested by Pauling [1] for the empirical radius for H in the bond distances, d(HX) of the partially ionic bonds in hydrogen halides (HX, where X = F, Cl, Br, I) is thus actually the Golden ratio based cationic radius, $d(H^+)$ of Eq. 12. Also, the ionic resonance forms [1] at the same equilibrium distance (0.74 Å) as d(HH) for the $H_2$ molecule are actually the cation ($H^+$) and anion ($H^-$) of H as in Eq. 12.

On subtracting d(H+) = 0.28 Å, from the experimental bond lengths d(HX) of hydrogen halides (HX) and d(MH) of alkali hydrides (MH), one obtains [7,8] the successive Eqs. (12,13):

$$d(HX) - d(H+) = d(X) = d(XX)/2; \qquad (\text{for } X = Cl, Br, I) \qquad (13)$$

$$d(MH) - d(H+) = d(M+) = d(MM)/\phi^2; \qquad (\text{for } M = Li, Na, K) \qquad (14)$$

where d(X) is found to be the covalent radius = d(XX)/2 of the halogens and d(M+) is found to be exactly = $d(MM)/\phi^2$ = Golden ratio based cationic radius of M and d(MM) is the inter-atomic distance of the edge atoms of the b.c.c. metal lattice [14].

The data [1, 13] on the bond distances d(AH) of some hydrides from [15] and the Golden ratio based radii of ions of many other atoms that account for the bond lengths d(AH) in hydrides are given in Table 1. The 1:1 correspondence of the radii sum with the observed bond lengths can be seen in Fig. 4. Many similar correlations can be found in [7]. Thus, bond lengths of completely or partially covalent or ionic bonds are sums of the radii of adjacent atoms or ions, where the latter have the Golden ratio based ionic radii. Therefore, in general, for any atom A,

$$d(AA) = 2d(A) = d(AA)/\phi^2 + d(AA)/\phi = d(A+) + d(A-) \quad (15)$$

where d(A) is the covalent radius and d(A+) and d(A-) are the Golden ratio based cationic and anionic radii of A.

From the data in Table 1, it can be seen that since the radius of hydrogen in hydrides has different values depending on the type of the atom or ion with which it combines, the recent article [16] providing an average value of 0.31 Å for the covalent radius for hydrogen and similar averages for other atoms can be erroneous.

## 4. ADDITIVITY OF THE GOLDEN RATIO BASED IONIC RADII IN ALAKALI HALIDES

On subtracting d(M+) of Eq. 14 from the known [14] inter-ionic distances d(MX) in alkali halides (MX), one finds that

$$d(MX) - \underline{d(M+)} = d(XX)/\phi = \underline{d(X-)}; \quad \text{(for MX, alkali halides)} \quad (16)$$

where $\underline{d(X-)} = d(XX)/\phi$, the Golden ratio based anionic radius of X and d(XX) is the covalent bond distance in the $X_2$ molecule [1]. These radii add up to give the exact crystallographic interionic distances [14] in the alkali halides [7]. See Fig. 5. Therefore, no radius ratio corrections as suggested in [1] are needed.

For the role of the Golden ratio and additivity of ionic radii in aqueous solutions and in the length of the hydrogen bonds, see [8a-c].

**Acknowledgement.** The author is grateful to the Institute of Biophysics of the Academy of Sciences of the Czech Republic (ASCR) for support by institutional research plans Nos. AV0Z50040507 and AV0Z50040702 grants of the ASCR.

## References

[1] Pauling L., "The Nature of the Chemical Bond" (Cornell Univ. Press, NY, 1960)

[2] http://wps.prenhall.com/wps/media/objects/3311/3390919/blb0702.html

[3] R. Heyrovska, a) Open Structural Biology J., **2,** 1-7 (2008); b) http://arxiv.org/abs/0708.1271

[4] http://arxiv.org/abs/0801.4261 (caffeine, etc) R.H. and S.N.

[5] http://arxiv.org/abs/0804.2488 amino acids

[6] http://arxiv.org/abs/0804.4086 Graphene, benzene,

[7] R. Heyrovska, Mol. Phys., **103,** 877-882 (2005).

[8] R. Heyrovska, a): *Innovations in ChemicalBiology*, edited by B. Sener, (Springer, New York, January 2009, Chapter 12), b) Chem. Phys. Lett., **429**, 600-605 (2006); c) Chem. Phys. Lett., **432**, 348-351 (2006).

[9] a) M. Livio, *The Golden Ratio, the story of phi, the world's most astonishing number*. (Broadway Books, New York, 2002); b) http://www.goldennumber.net;
c) http://en.wikipedia.org/wiki/File:Pentagon-construction.svg;
d) http://en.wikipedia.org/wiki/Pentagram;
e) http://mathworld.wolfram.com/RegularPolygon.html

[10] R. Heyrovska and S. Narayan, http://arxiv.org/abs/physics/0509207

[11] http://en.wikipedia.org/wiki/Luca_Pacioli (This year 2009 is the 500$^{th}$ Anniversary of: *De divina proportione* , written in Milan in 1496–98, published in Venice in 1509).

[12] a) P. J. Mohr and B. N. Taylor, Phys. Today, **54**, 29 (2001) and b)

http://physics.nist.gov/cuu/constants/

[13] E. A. Moelwyn-Hughes, *Physical Chemistry,* (Pergamon Press, London, 1957).

[14] C. Kittel, *Introduction to the Physics of Solids* (John-Wiley, New York, 1976).

[15] R. Heyrovska, 2004 International Joint meeting of ECS, USA and Japanese, Korean and Australian Societies, Honolulu, Hawaii, October 2004, Vol. 2004 - 2, Extended. Abs. C2-0551; http://www.electrochem.org/dl/ma/206/pdfs/0551.pdf

[16] B. Cordero, V. Gomez, A. E. Platero-Prats, M. Reves, J. Echeverria, E. Cremades, F. Barragin and S. Alvarez, Dalton Trans., 2832-2838 (2008).

**Table 1: Bond lengths d(AH) and radii of A and H in hydrides (in Å).**

| Atom A, [14] lattice, d(AA) | d(AA) [1,13,14] | Bond AH | d(AH) [1],[13] | Radius of H in d(AH) d(H) | Radius of A: d(A), d(A+) d(A) | d(AH),cal cols. 4+5 |
|---|---|---|---|---|---|---|
| diam,L*3$^{1/2}$/4= | 1.54 | **CH** | **1.10** | 0.37 | 0.77 | **1.14** |
| [1] | 1.40 | **NH** | **1.02** | 0.37 | 0.70 | **1.07** |
| [1] | 1.21 | **OH** | **0.96** | 0.37 | 0.61 | **0.98** |
| [1] | 2.20 | **PH** | **1.42** | 0.37 | 1.10 | **1.47** |
| [1] | 1.88 | **SH** | **1.34** | 0.37 | 0.94 | **1.31** |
| [1] | | | | **d(H)** | **d(A+)** | |
| [1] | 1.42 | **HF** | **0.92** | 0.37 | 0.54 | **0.91** |
| bcc, L= | 2.87 | **FeH** | **1.48** | 0.37 | 1.10 | **1.47** |
| hcp,2a/3$^{1/2}$= | 4.00 | **TlH** | **1.87** | 0.37 | 1.53 | **1.90** |
| | | | | **d(H+)** | **d(A)** | |
| [1] | 1.99 | **HCl** | **1.27** | 0.28 | 1.00 | **1.28** |
| [1],p.225 | 2.28 | **HBr** | **1.42** | 0.28 | 1.14 | **1.42** |
| [1],p.225 | 2.67 | **HI** | **1.61** | 0.28 | 1.34 | **1.62** |
| [1],p.368 | 1.77 | **BH** | **1.19** | 0.28 | 0.89 | **1.17** |
| fcc, a*0.707= | 2.86 | **AlH** | **1.65** | 0.28 | 1.43 | **1.71** |
| [1],p.224 | 2.08 | **SH** | **1.34** | 0.28 | 1.04 | **1.32** |
| hcp,a= | 2.50 | **CoH** | **1.54** | 0.28 | 1.25 | **1.53** |
| fcc, a*0.707= | 2.49 | **NiH** | **1.48** | 0.28 | 1.25 | **1.53** |
| [1],p.224 | 2.34 | **Se H** | **1.47** | 0.28 | 1.17 | **1.45** |
| hcp,a= | 2.66 | **ZnH** | **1.60** | 0.28 | 1.33 | **1.61** |
| hcp,a= | 2.98 | **CdH** | **1.76** | 0.28 | 1.49 | **1.77** |
| trig,a= | 3.01 | **HgH** | **1.74** | 0.28 | 1.51 | **1.79** |
| diam, L3$^{1/2}$/4= | 2.91 | **SnH** | **1.70** | 0.28 | 1.46 | **1.74** |
| diam, L3$^{1/2}$/4= | 2.35 | **SiH** | **1.48** | 0.28 | 1.18 | **1.46** |
| diam, L3$^{1/2}$/4= | 2.45 | **GeH** | **1.53** | 0.28 | 1.23 | **1.51** |
| trig,a= | 2.91 | **SbH** | **1.71** | 0.28 | 1.46 | **1.74** |
| | | | | **d(H+)** | **d(A+)** | |
| bcc, L= | 3.49 | **LiH** | **1.60** | 0.28 | 1.33 | **1.61** |
| bcc, L= | 4.23 | **NaH** | **1.89** | 0.28 | 1.61 | **1.89** |
| bcc, L= | 5.23 | **KH** | **2.24** | 0.28 | 2.00 | **2.28** |
| bcc, L= | 5.59 | **RbH** | **2.37** | 0.28 | 2.13 | **2.41** |
| bcc, L= | 5.02 | **BaH** | **2.23** | 0.28 | 1.92 | **2.20** |
| trig,a= | 3.16 | **AsH** | **1.52** | 0.28 | 1.21 | **1.49** |
| | | | | **d(H-)** | **d(A+)** | |
| bcc,L*0.866= | 5.23 | **CsH** | **2.49** | 0.46 | 2.00 | **2.46** |
| hcp,a= | 2.27 | **BeH** | **1.34** | 0.46 | 0.87 | **1.33** |
| hcp,a= | 3.21 | **Mg** | **1.73** | 0.46 | 1.23 | **1.69** |
| fcc, a*0.707= | 3.95 | **CaH** | **2.00** | 0.46 | 1.51 | **1.97** |
| fcc,a*0.707= | 2.55 | **CuH** | **1.46** | 0.46 | 0.97 | **1.43** |
| fcc,a*0.707= | 2.89 | **AgH** | **1.62** | 0.46 | 1.10 | **1.56** |
| fcc,a*0.707= | 2.88 | **AuH** | **1.52** | 0.46 | 1.10 | **1.56** |
| fcc,a*0.707= | 3.50 | **PbH** | **1.84** | 0.46 | 1.34 | **1.80** |
| fcc, a*0.707= | 4.30 | **SrH** | **2.15** | 0.46 | 1.64 | **2.10** |
| | | | | **d(H)** | **d(A-)** | |
| cub,a=2.24 | 2.24 | **MnH** | **1.73** | 0.37 | 1.38 | **1.75** |
| | | | | **d(H-)** | **d(A)** | |
| [13] | 2.80 | **InH** | **1.84** | 0.46 | 1.40 | **1.86** |

**(1a): <u>AB</u>: <u>any line.</u>**
AP & BP: **Golden sections, P: Golden point**
BC = **AB**/2.
AC = $5^{1/2}$**AB/2.**
AP = AD = **<u>AB/ϕ.</u>**
BP = AP/ϕ = **AB**/$\phi^2$.

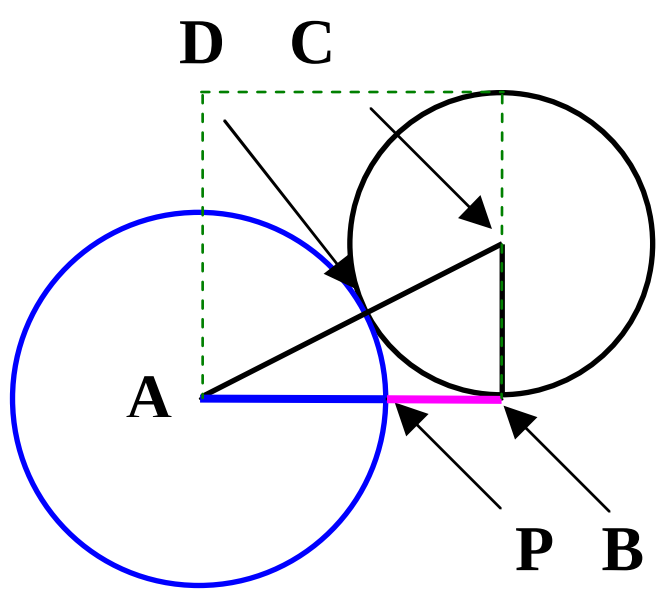

**(1b): <u>AB:</u>** <u>radius of a circle</u> as shown.
AP & BP: **Golden sections**; P: **Golden point**.

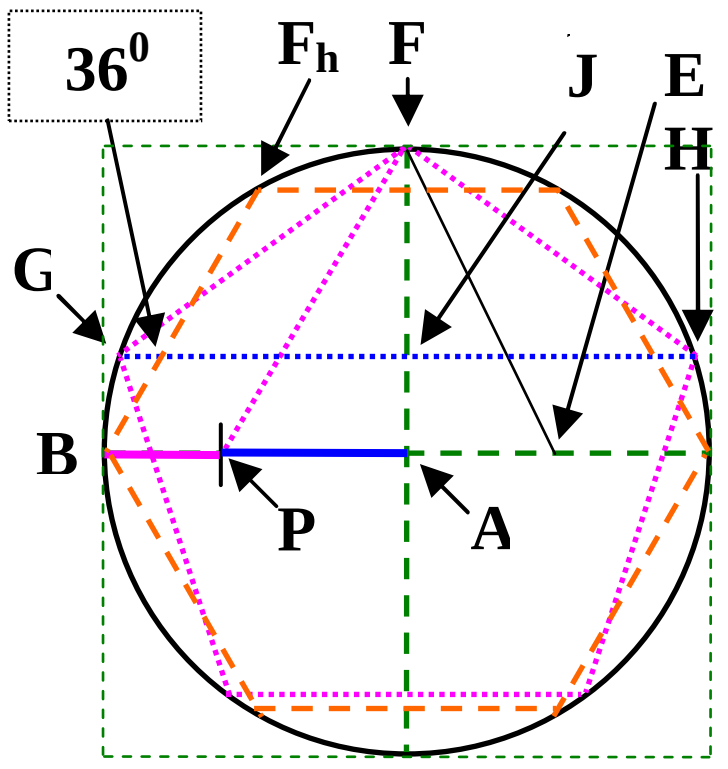

**AB** = AF = $BF_h$ **=** side of the hexagon and of a square.
AE = **AB**/2 .
EF = $5^{1/2}$**AB/2** = EP.
AP = **<u>AB/ϕ.</u>**
BP = AP/ϕ = **AB**/$\phi^2$.
FP = FG = FH =_ side of the pentagon
= GH/ϕ =1.176**AB**
= $(1+1/\phi^2)^{1/2}$**AB.**
$\cos 36^0$ = GJ/FG = ϕ/2.

**(1c): The Golden ratio** in the pentagon, pentagram and decagon; **AB** = radius of the circumscribing circle.

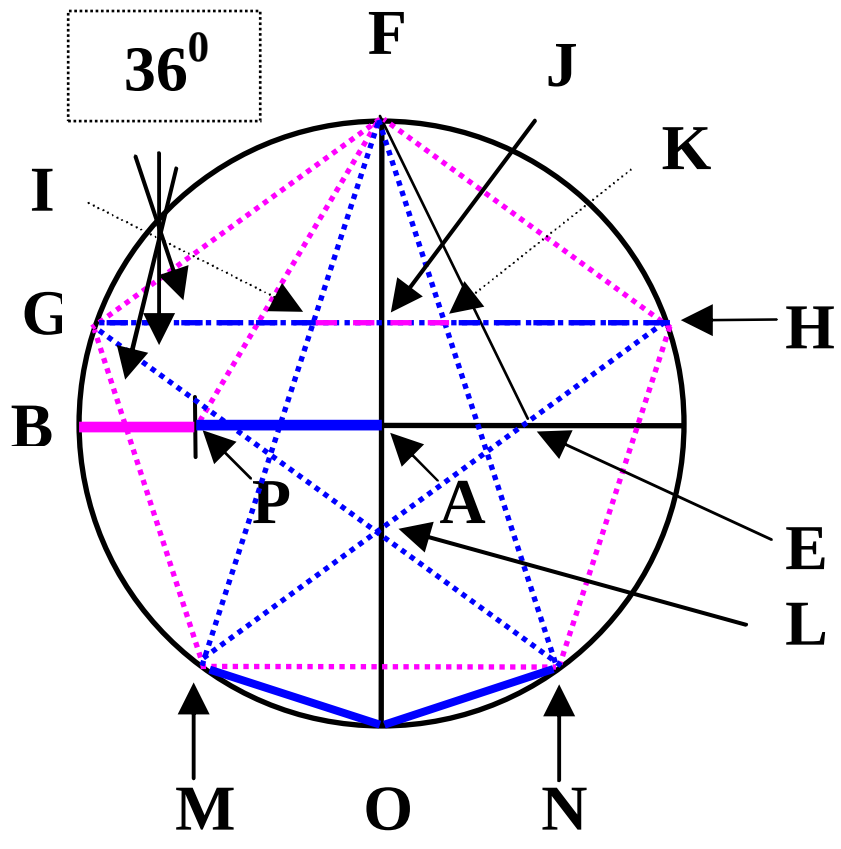

**AB**/AP = AP/BP = ϕ =
GH/FP = GH/FG =
GK/GI = GI/IK = FM/MN.
FG = MN = 2**AB**sin $36^0$ = 1.176**AB**
(= side of the pentagon; $36^0$ = π/5)
GJ/FG = GJ/GL = $\cos 36^0$ = ϕ/2.
sin (MNO) = $\sin 18^0$ = 1/2ϕ
MO = ON = MN/2$\cos 18^0$ =
2**AB**$\sin 18^0$ = **AB**/ϕ = AP
(= side of the decagon **= AP, the Golden section of AB:**

**<u>Fig. 1</u> The Golden point (P) and the Golden sections (AP & BP) of a line AB, [9a-e], (1a), [9b]:** The line BC = AB/2 is drawn perpendicular to AB and C is joined to A. The point D is marked on AC such that BC = CD. The <u>Golden point</u> P is marked on AB such that AP = AD. (**1b**), **[9c]:** AE = AB/2 and E is joined to F as shown. The <u>Golden point</u> P is marked on AB such that EF = EP. The points G and H are such that FP = FG = FH = GH//ϕ, the sides of a regular pentagon. AB is also the side of a square and of the regular hexagon. **(1c), [9d,e]** The pentagon, pentagram and decagon in the circle with AB as the radius.

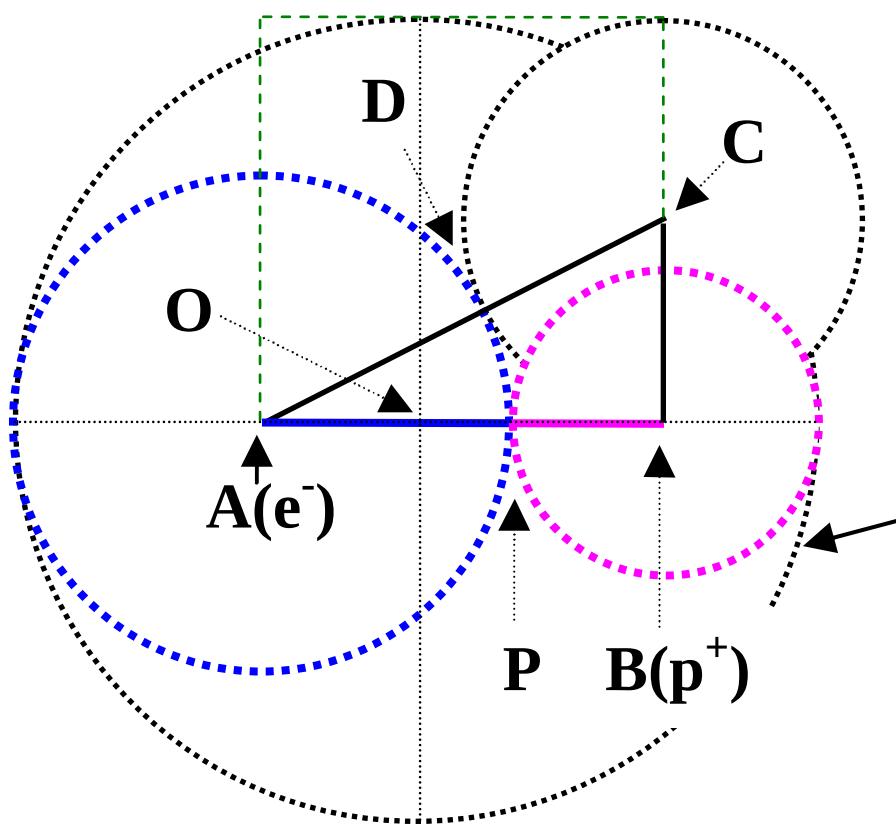

**Golden sections of the Bohr radius ($a_B$)**

**AB** = AP + BP = $a_B$ = 0.529Å.
BC = AB/2; AC = $5^{1/2}$AB/2.
AD = AP = $a_B$ = $a_B/\phi$ = 0.327Å.
BP = $a_B$ = $a_B/\phi^2$ = 0.202Å.
Radius of this circle = $a_B$; de Broglie wavelength, $\lambda_{dB} = 2\pi a_B$.
OP = AP – BP = $a_B/\phi^3$ = 0.125Å.
P: **Golden point**.

**Fig. 2:** Bohr radius, AB **=** $a_B$, Golden point P and Golden sections, AP and BP.

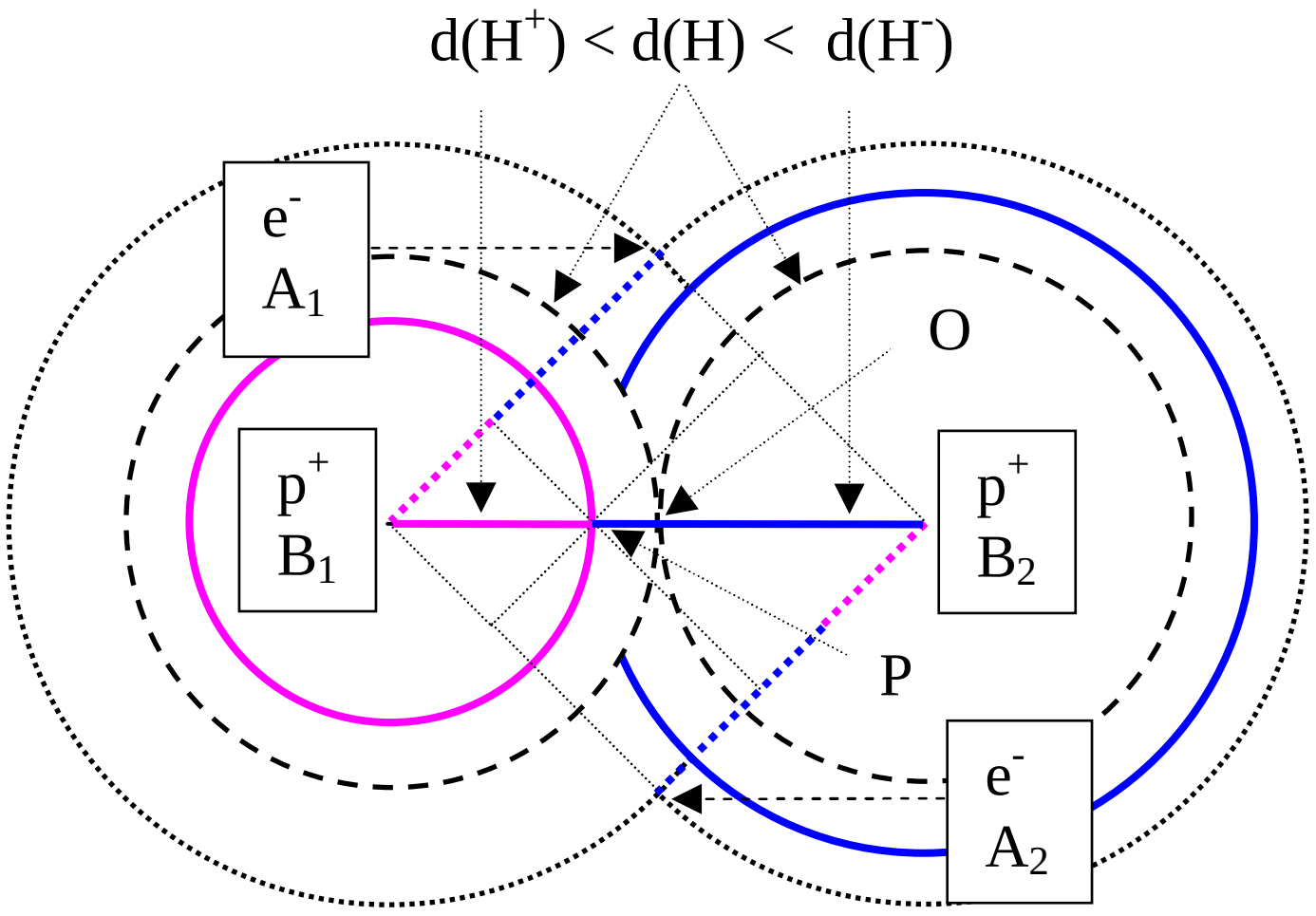

**Golden sections of the H-H bond length in $H_2$:**
$2^{1/2}a_B$ = d(HH) = d($H^+$) + d($H^-$)

$A_1B_1 = A_2B_2 = a_B$ (see Fig. 2)
$B_1B_2$ = d(HH) = $2^{1/2}a_B$ = 0.74Å
**$B_2P/ B_1P = B_1B_2/B_2P = \phi$.**
d(H) = $B_1O = B_2O$ = d(HH)/2
d($H^+$) = $B_1P$ = d(HH)/$\phi^2$ = 0.28Å
d($H^-$) = $B_2P$ = d(HH)/$\phi$ = 0.46Å
OP = $B_2P - B_2O = B_1O - B_1P$ = d(HH)(1/$\phi$ - ½) = d(HH)(½ - 1/$\phi^2$) = 0.118d(HH) = 0.087Å.

**Fig. 3:** The Golden ratio based radii of the H atom, d($H^+$) = $B_1P$ = d(HH)/$\phi^2$ and d($H^-$) = $B_2P$ = d(HH)/$\phi$ of hydrogen. P is the Golden point on d(HH) = $B_1B_2$.(= 0.74 Å [1], 0.75 Å [13]). The circles with radii $A_1B_1 = A_2B_2 = a_B$ intersect at $A_1$ and $A_2$, and the two electrons are shared by the two protons of the $H_2$ molecule.

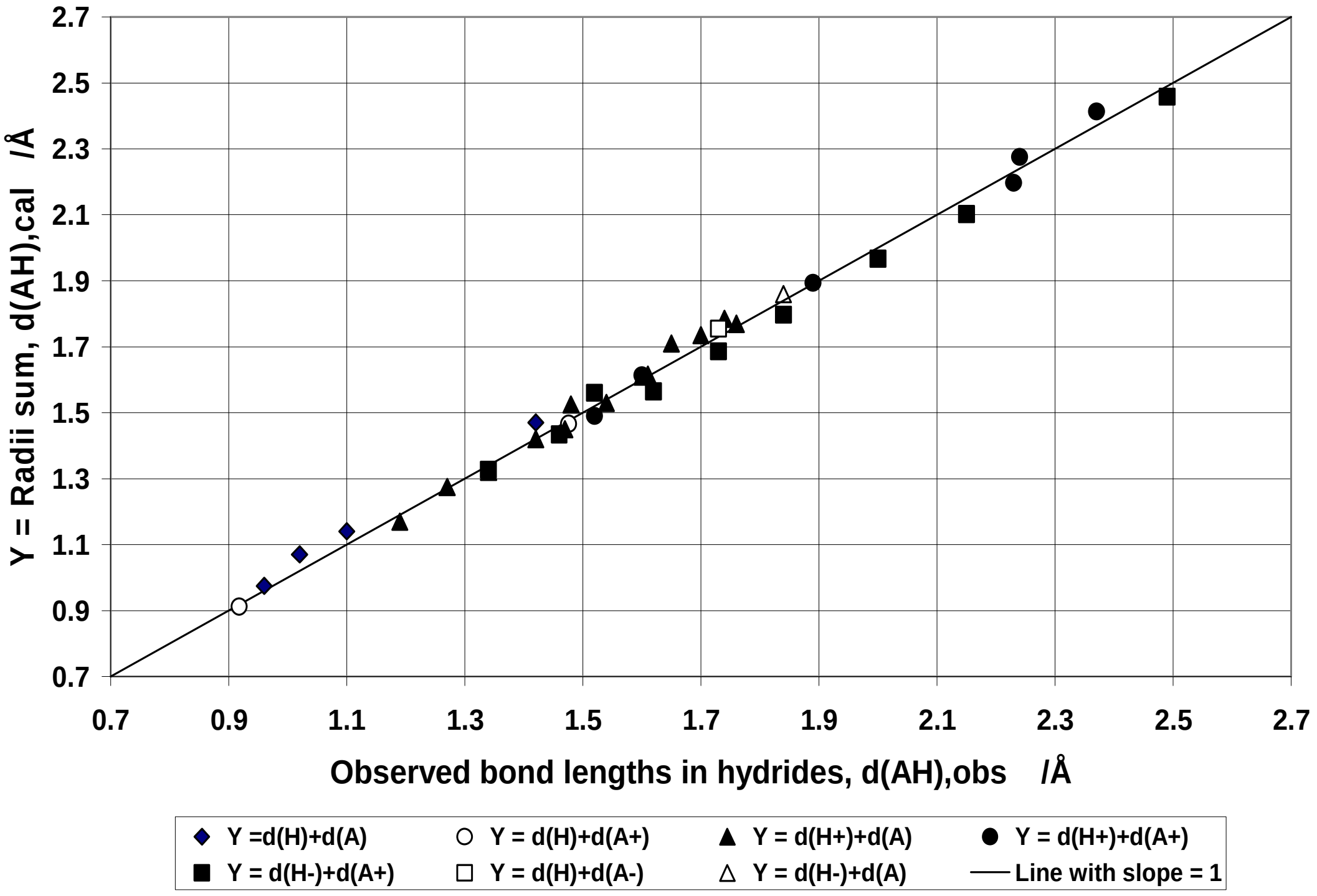

**Fig. 4.** Linear graph of the radii sum versus the observed bond lengths for hydrides.

## The Golden ratio based ionic radii

**(a) Cationic radii of Group I elements:** $d(A+) = d(AA)/\phi^2 < d(AA)/2$

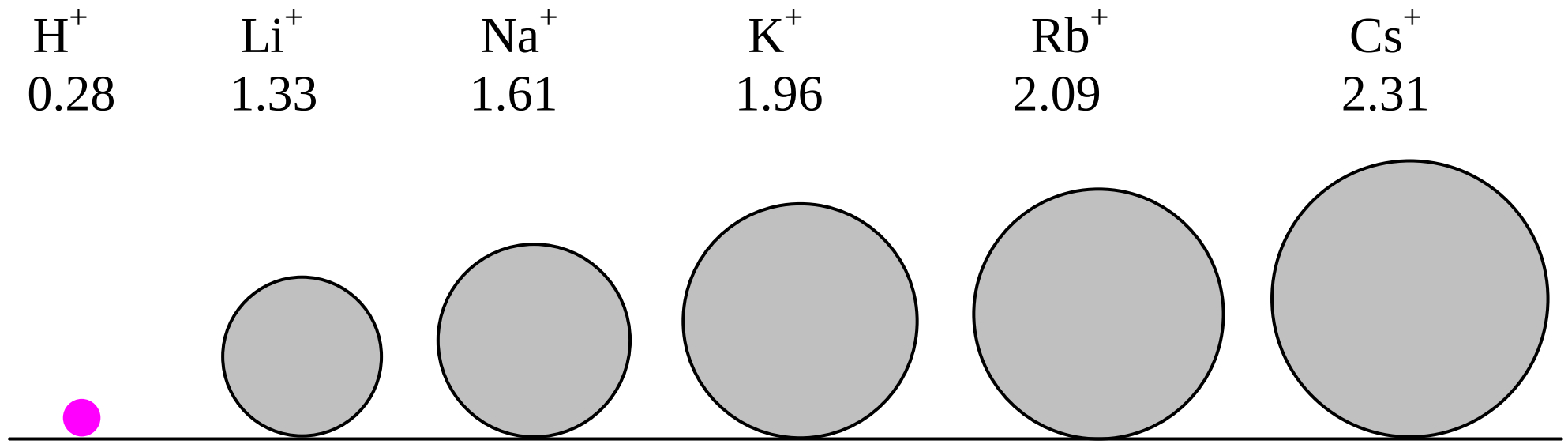

**(b) Anionic radii of Group VII elements and $H^-$:** $d(A-) = d(AA/\phi) > d(AA)/2$

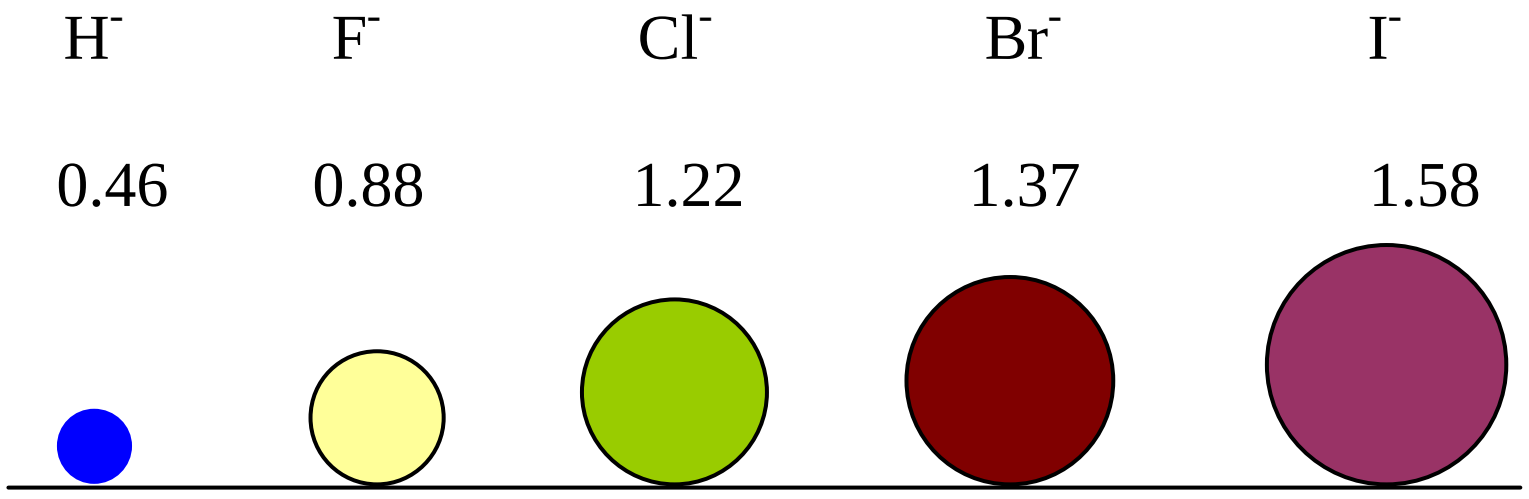

**(c) Inter-ionic distances in alkali halides = sum of the ionic radii:**

Example: $d(NaCl) = d(Na^+) + d(Cl^-) = 2.83$ Å.

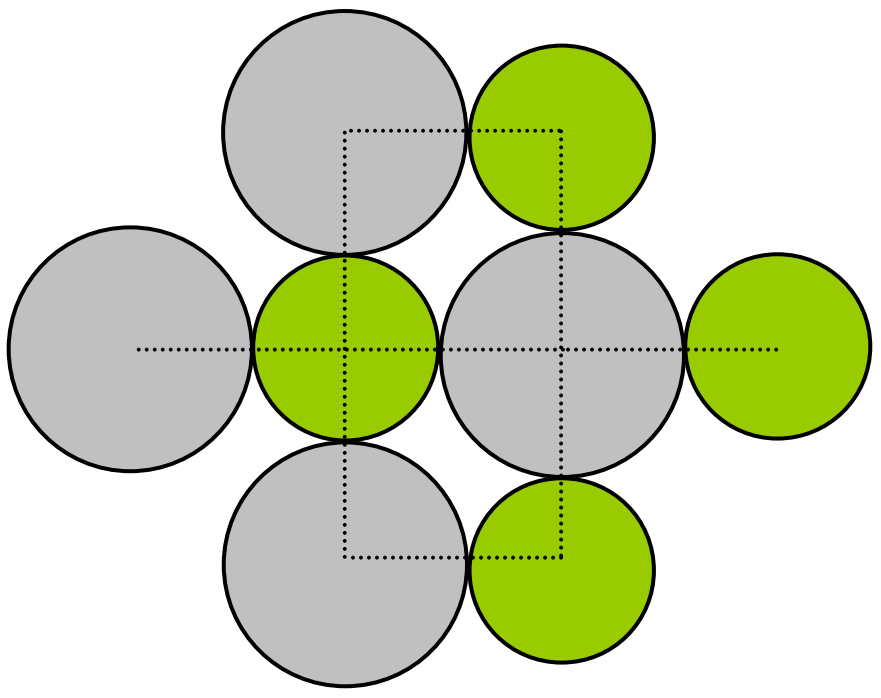

**Fig. 5, 1a-c: (a)** Alkali metal cations, **(b)** halide anions and their Golden ratio based ionic radii in (Å). The hydrogen cation and anion are also given for comparison of the sizes. **(c)** The face centered NaCl showing the additivity of the ionic radii: $d(NaCl) = d(Na^+) + d(Cl^-) = 1.61 + 1.22 = 2.83$ Å [7], (observed value = 2.82 Å, [14]). The additivity of radii holds for all the alkali halides and hydrides [7].